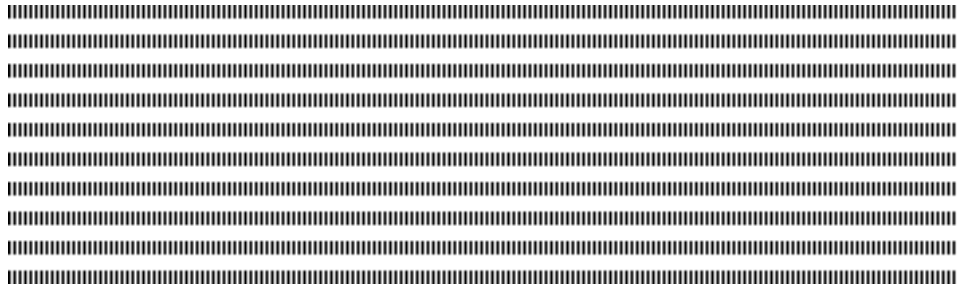

# DEVELOPPEMENT DE METHODES AUTOMATIQUES POUR LA REUTILISATION DES COMPOSANTS LOGICIELS

KOFFI Kouakou Ive Arsène, Docteur BROU Konan Marcellin, Prof. OUMTAGADA Souleymane

Département d'informatique
Ecole Doctorale Polytechnique
Institut National Polytechnique
Houphouet Boigny (EDP-INPHB)
YAMOUSSOUKRO
CÔTE D'IVOIRE
koffiyvesarsene@gmail.com

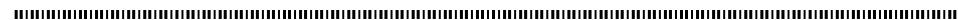

**RÉSUMÉ.** La masse importante d'informations et à la complexité croissante des applications contraignent les développeurs à avoir des composants autonomes et réutilisables des bibliothèques et des marchés de composants. Notre approche consiste à développer des méthodes pour évaluer la qualité du composant logiciel de ces bibliothèques d'une part et d'autre part à optimiser le coût financier et le temps d'adaptation des composants sélectionnés. Notre fonction objectif définit une métrique qui maximise la valeur de la qualité du composant logiciel en minimisant le coût financier et le temps de maintenance. Ce modèle devrait permettre de classer les composants et de les ordonner afin de choisir le plus optimisé.

**ABSTRACT.** The large amount of information and the increasing complexity of applications constrain developers to have stand-alone and reusable components from libraries and component markets. Our approach consists in developing methods to evaluate the quality of the software component of these libraries, on the one hand and moreover to optimize the financial cost and the adaptation's time of these selected components. Our objective function defines a metric that maximizes the value of the software component quality by minimizing the financial cost and maintenance time. This model should make it possible to classify the components and order them in order to choose the most optimized.

**MOTS-CLES :** développement de méthode, réutilisation, composants logiciels, qualité de composant

**KEYWORDS:** method development, reuse, software components, component quality
.

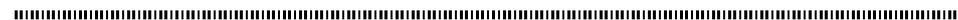

**INTRODUCTION**

Les développeurs et les entreprises se trouvent souvent confrontées à une masse importante d'informations dans l'ingénierie des systèmes d'information. Cette masse importante d'informations a pour conséquence l'augmentation de la taille des logiciels à développer et l'accroissement de la complexité de ces applications. Pour résoudre ces difficultés, les développeurs de logiciel font de plus en plus recours à des composants réutilisables dans leurs applications.

La réutilisation de ces composants nécessite le développement de modèles et de méthodes sans lesquels de nombreuses tâches sont manuelles et répétitives.

Nous allons, dans ces travaux de recherche, développer des méthodes automatiques pour évaluer et améliorer la qualité des composants logiciels sélectionnés selon les critères définis par l'utilisateur.

Il s'agit de :
1. Définir un modèle de qualité du composant logiciel;
2. Définir un processus de sélection du composant pertinent et adapté au système logiciel à construire;
3. Etablir un modèle métrique pour évaluer, maximiser la qualité du composant.
4. Optimiser le coût et le temps de maintenance.

Ce travail est organisé comme suit. La première partie concerne l'état de l'art relatif à la sélection des composants réutilisables. La deuxième partie traite des différents modèles développés qui font objet de notre article. La dernière partie concerne la conclusion et les perspectives.

**SECTION 1**

# 1. Etat de l'art
## 1.1. méthodes d'analyse multicritère et optimisation

Pour évaluer la qualité des objets sélectionnés en général et des composants logiciels en particulier, plusieurs méthodes de prise de décision multicritère ont été réalisées.

Dans l'article de E. Triantaphyllou et al.(1998), les chercheurs proposent la méthode d'analyse hiérarchique des procédés (AHP). Cette technique établit une table de comparaison binaire des caractéristiques, des sous caractéristiques des composants logiciels par niveau. Ensuite détermine les poids de ces différents critères et sous critères d'une part et d'autre part évalue la cohérence de cette table.

Dans les travaux de A.A. Zaidan et al.(2015), les auteurs définissent une approche de prise de décision multi-critères pour traiter les problèmes complexes. Cette approche permet de décomposer le problème en plusieurs niveaux, définissant les objectifs et fournissant un cadre d'ensemble pour l'évaluation de solutions. Ils proposent la méthode d'analyse AHP comme la meilleure alternative pour la maximisation des données complexes. Ils donnent le modèle suivant :

$$A_{AHP}^* = \max \sum_{j=1} q_{ij} w_j \; for \; i = 1, 2, 3, ....M \tag{1}$$

Dans le cadre de l'optimisation des paramètres liés aux composants sélectionnés, l'article de R. Perriot et al.(2014), donne différents modèles mathématiques d'optimisation en programmation linéaire. L'un de ces modèles établit un compromis entre le coût monétaire minimum et le temps de réponse dans les nuages informatiques. Il est formulé ci-dessous :

$$\begin{cases} \text{minimiser}(\alpha C + (1-\alpha)T) \\ \text{avec les contraintes } qui\,definissent \\ C: le \text{ modèle de cout} \\ T: le\,temps\,de\,selection\,des\,vues \end{cases} \tag{2}$$

## 1.2. Sélection de composants logiciels

La gestion des informations implique la recherche, la sélection et le stockage de documents pertinents en général et dans le domaine du génie logiciel, des composants logiciels en particulier.

Le chercheur E. Rames (1991) définit un modèle de recherche basé sur la classification hiérarchique et thématique des composants logiciels contenus dans une base. Cependant, sa méthode de classification des composants était basée sur une technique manuelle. Dans l'article de B. George et al.(2010), un mécanisme permettant l'automatisation de la sélection d'un composant logiciel, parmi un ensemble de candidats en fonction de leurs propriétés fonctionnelles et non fonctionnelles a été étudié.

Ce mécanisme permet de rendre possible l'extraction et de comparer les composants. Il s'agit après des phases de sélection de composants, *de mesurer l'indice de satisfaction* de ces différents composants candidats sélectionnés afin de trouver les plus pertinents.

L'article de A.A. Zaidan et al.(2015) propose une étude comparative des logiciels dont le but est d'évaluer et de sélectionner des logiciels « open source » pour la gestiondes dossiers médicaux électroniques et numériques. Cette étude est réalisée avec différentes techniques de prise de décision à critères multiples. Ainsi les systèmes logiciels sont sélectionnés sur la base d'un ensemble de résultats métriques à l'aide de la technique AHP intégrée à différentes techniques de prise de décision multicritères.

Dans l'article de J. Pande et al.(2013), une métrique de *souplesse* a étédéfinie. Cette métrique leur a permis de déterminer une sélection optimale des composants logiciels avec le modèle suivant :

$$p = \sum_{h \in A} \sum_{i \in SC} w_h q_{hi} x_i - \sum_{i \in SC} c_i x_i \qquad (3)$$

A = ensemble d'attributs de qualité ;

SC = ensemble de composants disponibles (composants candidats) ;

$q_{hi}$ = le niveau normalisé de l'attribut de qualité h∈ A pour le composant i ;

$W_h$ = poids attribué à l'attribut de qualité h∈ A ;

$x_i$ = 1 si le composant *i* est sélectionné, sinon 0 ;

$C_i$ = coût normalisé du composant *i*.

### 1.3. Limite des méthodes

Compte tenu des méthodes développées, nous remarquons que différents chercheurs ont apporté d'énormes contributions. Cependant, certains aspects tels que le temps d'adaptation et de maintenance des composants logiciels d'une part et d'autre part, l'optimisation des deux paramètres coût et temps relatifs aux composants sélectionnésde ces bibliothèques n'ont pas été pris en compte.

**SECTION 2**

## 2. Problème de recherche

Dans la revue de littérature, divers travaux relatifs aux méthodes de sélection de composants logiciels pertinents ont été réalisés.

Les travaux que nous présentons traitent de la problématique de l'évaluation de la qualité des composants pré-faits. Il s'agit de la maximisation de leurs valeurs de qualité en optimisant le coût financier, de maintenance et le temps d'adaptation de ces composants en vue de les réutiliser dans un système logiciel.

Notre objectif est donc de déterminer une métrique qui permettra de maximiser la qualité du composant logiciel sélectionné tout en minimisant le coût financier, de maintenance et le temps d'adaptation de ce composant.

Il s'agit alors de définir un modèle de qualité du composant logiciel, ensuite de définir un processus de sélection des composantsadaptés au système logiciel à construire, et enfin d'établir un modèle métrique pour évaluer la qualité du composant logiciel sélectionné d'une part, et d'autre part, d'optimiser le coût financier et le temps de modification, d'adaptation de ce composant.

Ceci nous amène à formuler les hypothèses de recherche suivantes :

Les caractéristiques définies des composants logiciels permettent-elles d'obtenir des attributs mesurables et générant des composants logiciels de qualité ?

La sélection de composant de qualité, pertinent répondant aux besoins de l'utilisateur permet-elle d'optimiser le coût financier de maintenance et le temps d'adaptation de ce composant dans un système logiciel?

**SECTION 3**

## 3. Phase de modélisation
### 3.1. Proposition de modèle de qualité du composant logiciel

Nous nous intéresserons à l'évaluation de la sélection et à l'intégration des composants logiciels dans un système de logiciel. Notre objectif principal est de faire le choix du « meilleur composant logiciel » d'une bibliothèque ou d'un marché de composants. Cette sélection doit répondre au mieux aux critères définis par l'utilisateur et selon le type d'application à construire. Les caractéristiques définies sont les suivantes : la capacité fonctionnelle, la fiabilité, la facilité d'utilisation, la sécurité et la maintenabilité.

Ce qui nous permet de définir le modèle[1] suivant :

*Fig1 modèle de qualité de composant logiciel*

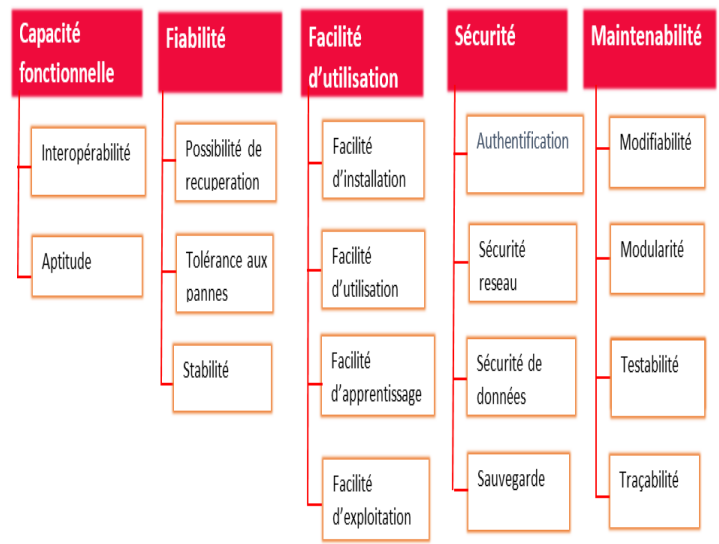

Ce modèle est basé sur le modèle de qualité ISO 9126 et des représentations de qualité des revues de littérature. Il permet de préciser les caractéristiques les plus importantes pour le choix des composants logiciels selon les besoins de l'utilisateur.

---

[1]*Modèle de qualité, Inspiré du modèle ISO 9126 et de la qualité logicielle définie par Jérémie Grodziski*

ARIMA - Phase de Modélisation

En utilisant la technique Analyse Hiérarchique des Procédés (AHP), nous pourrons atteindre l'objectif qui consiste à choisir le composant logiciel qui répond le mieux possible aux besoins de l'utilisateur. Pour se faire, nous avons construit le modèle hiérarchique de qualité en fonction des caractéristiques et des sous caractéristiques des composants logiciels (fig1). Ensuite, à l'aide de la méthode d'analyse multicritère, nous avons construit une table de comparaison binaire des caractéristiques et sous caractéristiques. Ceci a permis de déterminer les poids des différents critères de qualité définis du composant logiciel d'une part et d'autre part d'évaluer la cohérence de notre travail.

### 3.2. Proposition de processus de sélection du composant logiciel

Nous allons donner une description du processus de sélection des composants sélectionnés puis évalués. Ceci nous amènera à faire le choix du composant pertinent et optimisé.

*Fig2 Modèle d'évaluation de la qualité du composant logiciel*

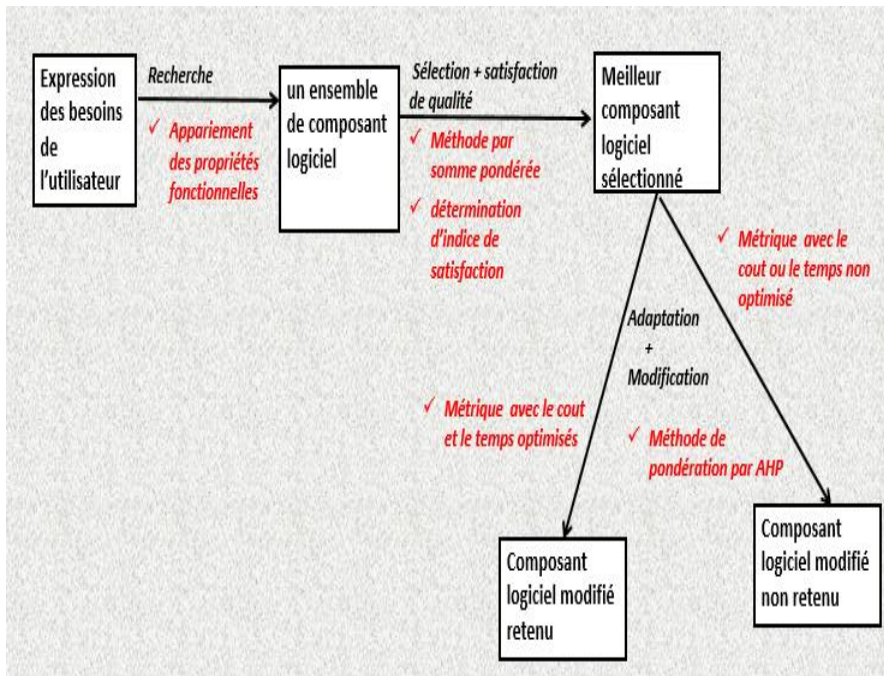

*En noir : les actions à effectuerlors de la sélection*
*En rouge : les méthodes utilisées lors de la sélection*

Ce processus est modélisé en UML de la manière suivante :

*Fig3   processus de sélection d'un composant*

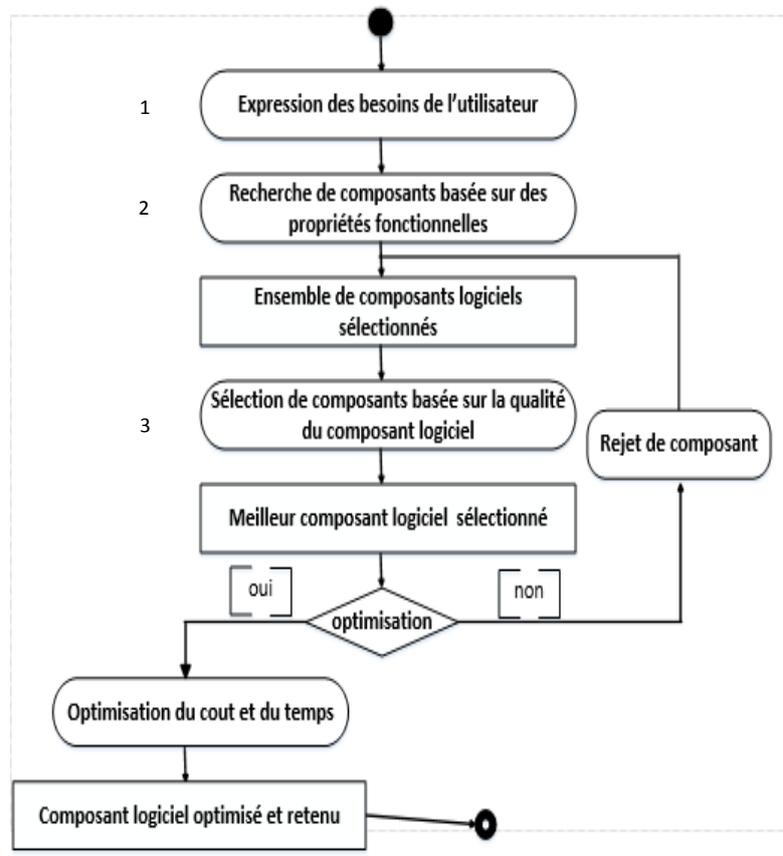

Il s'agit de définir un processus de sélection qui permettra à l'utilisateur de faire le choix du composant logiciel dans une bibliothèque. En effet sur le net, il existe de nombreuses bibliothèques telles que ComponentSource, Sourceforge, *flashline,citerAlterWay*, etc,  permettant de programmer une interface graphique pour des applications données.

Le processus de sélection suit les étapes suivantes :
**Etape1:** L'utilisateur exprime les besoins de qualité du composant.
**Etape2:** Une première recherche consiste à prendre en compte les propriétés fonctionnelles fournissant certains services en rapport avec le type de logiciel à construire et surtout les besoins exprimés par l'utilisateur.
Nous obtenons un ensemble de composants logiciels sélectionnés dont les propriétés sont des propriétés fonctionnelles fournies par les composants logiciels. Autrement dit, il s'agit des services rendus par les différents composants.
**Etape3:** Cette étape consiste à faire une sélection basée sur les propriétés non fonctionnelles. Il s'agit de tenir compte de la qualité du composant logiciel et comment les services sont rendus. Cette étape consiste à évaluer la qualité du composant à partir de métrique définie. On sélectionnera le composant qui répond aux mieux, aux critères de qualité définis par l'utilisateur.
**Etape4:** Au niveau de cette étape, on évalue la maintenance c'est-à-dire, la phase de modification et d'adaptation du composant dans un système en cours d'utilisation. Elle représente la phase de simulation pour déterminer la qualité du composant.
Nous appliquons à ce niveau, la métrique qui permet d'évaluer le coût financier et le temps générés. Elle évalueet produit un coût financier et un temps de maintenance.
**Etape5:** Dans le cas où les paramètrescoût et temps sont optimisés, alors, le composant sélectionné est retenu.
**Etape6:** Si les paramètres ne le sont pas, alors la recherche continue et le processus reprend

### 3.3. Proposition de modèle pour évaluer le coût financier et le temps de maintenance

En nous inspirant des modèles définis dans les revues de littératures, notre approche consiste à évaluer en plus du coût financier du composant, le *temps* mis pour réaliser la maintenance.

Notre modèle a pour objectif de prendre en compte le paramètre *temps* lors de d'évaluation de la qualité du composant logiciel à l'aide de la programmation linéaire. Il s'agira pour nous de définir une métrique à deux paramètres dont le coût financier et le temps sont fonction de l'indice du composant choisi. Cette métrique sert à optimiser les paramètres afin de sélectionner le « meilleur composant logiciel » d'une part et d'autre part de *minimiser* ce coût financier et le temps d'adaptation et de modification du composant sélectionné.

Nous définissons notre fonction de la manière suivante :

$$f(c,t,i) = \alpha C_i + (1-\alpha) t_i \quad \forall i \in sc ; \tag{4}$$

Avec les contraintes suivantes :

$$0 \leq \alpha \leq 1$$

ARIMA -  Phase de Modélisation

$$t_i = \frac{t}{T_{\max}} \text{ et } \quad 0 \leq t_i \leq 1$$

$$c_i = \frac{c}{C_{\max}} \text{ et } \quad 0 \leq c_i \leq 1$$

Où

Sc : ensemble de composants disponibles ;
$C_i$ : Coût financier normalisé de maintenance du composant $i$ ;
$C$ : cout relatif généré par le composant $i$ ;
$C_{max}$ : Coût maximum réalisé par un des composants sélectionnés ;
$t_i$ : temps d'adaptation et de maintenance normalisé du composant $i$ ;
$t$ : temps relatif, généré par le composant $i$ ;
$\alpha$ : Coefficient d'adaptation ;
$T_{max}$ : temps maximum réalisé par un des composants sélectionnés.

Donc la métrique pour tout composant logiciel *i* sélectionné sera :

$$S_i = \sum_{h \in A} w_h q_{hi} x_i - \left[\alpha c_i + (1-\alpha) t_i\right] x_i \quad \forall i \in SC \quad \text{Et les contraintes} \tag{5}$$

et $\quad t_i = \dfrac{t}{T_{\max}} \quad 0 \leq t_i \leq 1 \quad\quad 0 \leq \alpha \leq 1$

$c_i = \dfrac{c}{C_{\max}} \quad\quad$ et $\quad 0 \leq c_i \leq 1$

Où
A = ensemble des caractéristiques de qualité du logiciel;
SC = ensemble de composants disponibles (composants candidats) ;
$q_{hi}$= le niveau normalisé de l'attribut de qualité h∈ A pour le composant i ;
$W_h$ =  poids attribué à l'attribut de qualité h∈ A ;
$x_i$ = 1 si le composant *i* est sélectionné, 0 sinon ;
$C_i$ = coût normalisé du composant *i* ;
$t_i$ : Temps normalisé de maintenance du composant $i$ ;
$\alpha$ : Coefficient d'adaptation à préciser

Le modèle (5) représente la fonction objectif. Cette fonction permet de calculer puisd'évaluer la qualité des caractéristiques des composants logiciels sélectionnés.

ARIMA - Phase de Modélisation

Pour maximiser la fonction objectif, nous devons optimiser les paramètres Temps et coût de maintenance. Alors, il nous faudrait maximiser le premier terme et minimiser le second terme de $S_i$.

Ceci revient à maximiser le terme $Q_i$

$$Q_i = \sum_{h \in A} w_h q_h x_i \quad \forall i \in SC \qquad (6)$$

et à minimiser le second terme $m_i$

$$m_i = \left[\alpha c_i + (1-\alpha) t_i\right] x_i \quad \forall i \in SC \qquad (7)$$

Pour tout composant logiciel $i$ de la bibliothèque, on obtient le système suivant :

$$\begin{cases} \max(\sum_{h \in A} w_h q_{hi} x_i - \left[\alpha c_i + (1-\alpha) t_i\right] x_i) \\ \sum_{h \in A} w_h = 1 \\ 0 \leq \alpha \leq 1 \\ t_i = \dfrac{t}{T_{\max}} \; et \; 0 \leq t \leq T_{\max} \\ c_i = \dfrac{c}{C_{\max}} \; et \; 0 \leq c \leq C_{\max} \\ x = 1 \; si \, selectionné \, sinon \, x = 0 \end{cases} \Leftrightarrow \begin{cases} \max(\sum_{h \in A} w_h q_{hi} x_i) - \min(\left[\alpha c_i + (1-\alpha) t_i\right] x_i) \\ \sum_{h \in A} w_h = 1 \\ 0 \leq \alpha \leq 1 \\ t_i = \dfrac{t}{T_{\max}} \; et \; 0 \leq t \leq T_{\max} \\ c_i = \dfrac{c}{C_{\max}} \; et \; 0 \leq c \leq C_{\max} \\ x = 1 \; si \, selectionné \, sinon \, x = 0 \end{cases} \qquad (8)$$

Nous pourrons alors comparer et ordonner les différentes valeurs $S_i$ désignant les valeurs de qualité de chaque composant logiciel choisis. Ainsi plus la valeur est grande, plus le composant logiciel est de qualité avec un temps d'adaptation et un coût de maintenance réduits.

**SECTION 4**

## 4. Phase de Validation

En matière de recherche, toute théorie doit passer par une phase d'expérimentation ou de simulation avant sa validation. Pour se faire, nous proposons un algorithme pour soutenir et valider la théorie développée évaluant la qualité du composant logiciel. Il s'agit également d'optimiser les deux paramètres dont le temps et le coût d'adaptation.
En effet nous proposons l'algorithme SelectCompo pour résoudre le problème

### 4.1. Tableau d'analyse du problème

L'algorithme SelectCompo a pour but de sélectionner, dans une bibliothèque choisie, un ensemble de composants (Pi). Nous présentons dans le tableau 1, les variables et les notations utilisées dans nos algorithmes.

| Identificateurs | Types | Entrée/ Sortie | Rôles |
|---|---|---|---|
| biblio | Chaine de caractère | Entrée | Variable désignant l'intitulé de la bibliothèque courante. |
| P[], composant[ ] | Chaine de caractère | Entrée | Variable désignant un tableau de composants de la bibliothèque courante. |
| $M_{ax}C_{ompo}$ | Entier | Entrée | Variable désignant le nombre maximum de composants de la bibliothèque courante. |
| $M_{ax}C_{aracter}$ | Entier | Entrée | Variable désignant le nombre maximum de caractéristiques et sous caractéristiques du composant courant |
| i,j | Entier | Entrée | Indices de parcours des composants |
| compo | Chaine de caractère | sortie | Variable désignant le composant optimisé et retenu. |
| nom | Chaine de caractère | Entrée / Sortie | Variable désignant le nom de la caractéristique Et de la sous caractéristique |
| poids | réel | Entrée / Sortie | Variable désignant le poids de la caractéristique et de la sous caractéristique |
| note | réel | Entrée / Sortie | Variable désignant le note de la caractéristique et de la sous caractéristique |
| SousCaract[ ] | Chaine de caractère | Entrée / Sortie | Variable désignant un tableau de caractéristiques et de sous caractéristiques du composant courant |
| Somme[] | réel | Sortie | Variabledésignant un tableau de valeurs réelles exprimant la qualité des composantschoisis |

| $C_r[]$ | réel | Entrée | Variable désignant un tableau de cout relatif du composant courant |
|---|---|---|---|
| $T_r []$ | réel | Entrée | Variable désignant un tableau de temps d'adaptation relatif du composant courant |

Tableau 1

### 4.2. Procédures utilisées

Nous allons initialiser les 5 principales caractéristiques (fig1) définies par l'utilisateur pour faire notre étude. Après la phase de sélection de l'ensemble des composants (Pi), nous allons évaluer la qualité des caractéristiques de ses composants en fonction de leur importance (poids) et de la qualité relative du composant (note) à l'aide de la procédure **evaluation(Pi,nbre)**. A l'étape suivante, nous allons optimiser les composants en fonction du coût et du temps de d'adaptation par laprocédure **optimisation(Pi,nbre)**. Enfin nous allons choisir le composant le plus optimisé de l'ensemble des composants Pi avec la procédure **affichage**(). Ce dernier correspondra au composant ayant la valeur maximale des Pi notée Max,suite à l'application de la procédure optimisation(Pi,nbre). Alors pour écrire l'algorithme SelectCompo, nous allons utiliser différentes procédures suivantes :

a. **Procédure initialisation().**

Elle permet d'initialiser les caractéristiques et les sous caractéristiques définies par l'utilisateur. Toutes les valeurs décrites (fig1) sont initialisées.

b. **Procédure saisie (biblio : chaine, nbre :entier)**

Cette procédure permet de saisir la bibliothèque et le nombre de composants disponibles et répondant aux critères définis par l'utilisateur en entrée.En sortie, nous récupérons la somme des poids des différentes caractéristiques.

c. **Procedure evaluation (Composant : tableau [1..$M_{ax}C_{ompo}$] : chaine, nombre : entier)**

Cette procédure permet de calculer par la métrique des valeurs de qualité définie. Elle prend en entrée une liste de composants et le nombre de composant souhaité. A la suite des tests et des contrôles, elle retourne un vecteur de valeurs réelles désignant la valeur de qualité du composant notée Somme [].

d. **Procedure optimisation** (**nbre : entier)**

Cette procédure permet de classer et d'ordonner les valeurs de qualité retournées par la procédure*evaluation*. En paramètre, nous avons le nombre de composants retenus à trier.

    e. **affichage()**

Cette procédure permet d'affiche le composant retenu après le tri par ordre décroissant des valeurs de qualité des composants

### 4.3. Présentationdes algorithmes

Dans cette partie de notre travail, nous présenterons le pseudo code dans les différents algorithmes.

#### 4.3.1. Algorithme SelectCompo

Le programme principal va se présenter comme suit :

**Algorithme SelectCompo**
  **Entrée :** biblio,
        tableau_de_composants   chaine
        Nbre, i  entier
        Somme [] tableau de réels
**Sortie :** compo  chaine
Debut
**Tantque** (besoins Exprimés)
**Debut**
 **saisie** (biblio , nombre )
 nbre = nombre
 Pour i de 1 à nbre faire
 Debut
**Selectionner (Pi, nbre)**
Pi = tableau_de_composants
 Finpour
 **initialisation()**
**Fintanque**
**si** ((conditionsCaracterisques Remplies) et (coût et temps relatifs dans
    intervalles requis)  **alors**
   **debut**
Pour i de 1 à nbre faire
Debut
   **evaluation (Pi, nbre**)
   somme[i] = ValeurQualite(Pi)

    Finpour
**Si** SatisfactionQualité**alors**
**Optimisation** (nbre )
**Sinon**
    Reverser (composants dans biblio)
  **finsi**
Afichage()
Fin
*Fig 4 :  pseudo code de selectCompo*

### 4.3.2. Algorithme Evaluation

Nous avons donné la fonction objectif $S_i$ représentée par le modèle (5).Cette fonction permettra l'évaluation de la qualité du composant logiciel de la bibliothèque en tenant compte des contraintes définies au modèle (8).

*L'algorithmeEvaluation* permet de calculer les valeurs de qualité définie par $S_i$. Cette procédure prend en paramètre, une liste de composants et le nombre de composant obtenu après la phase de sélection basée sur les propriétés fonctionnelles de celui-ci. A la suite des tests et des contrôles, elle retourne un vecteur de valeurs réelles désignant la valeur de qualité des composants.

**procedure Evaluation**(Composant : tableau [1..$M_{ax}C_{ompo}$] :  chaine**,**nombre : entier)
    **Entrée** :i,j entier
        Som_qlte ,
Som_CoutTemps   réel
**Sortie** :somme somme  tableau [1.. $M_{ax}$] de réels
    debut
        Ecrire(le nombre de composants)
        Lire(nombre)
         Si (nomnbre>$M_{ax}C_{ompo}$) alors
           nombre = $M_{ax}C_{ompo}$
      pour i de 1 à nombre
      debut
       Ecrire(cout et temps relatifs au composant  courant)
lire($c_r[i]$, $T_r[i]$))
Som_qlte ←0 ;
        Som_coutTemps←0 ;
         Pour j de 1 à $M_{ax}C_{aracter}$ faire
          Debut

Si (($0 \leq$ Caract[j].poids$\leq 1$) et ($0 <$ Caract[j].note $\leq Q_{max}$) et Somm =1 )  alors
　　Debutsi
$$W[j] \leftarrow caract[j].poids$$
$$q_r[j] \leftarrow caract[j].note$$
$$q[j] \leftarrow \frac{q_r[j]}{Q_{max}}$$
　　Som _qlte ← Som _qlte + w[j]*q[j]
　　Finsi
　Finpour
Si ( $0 < T_r[i]) \leq T_{max}$) et ( $0 < C_r[i]) \leq C_{max}$)  alors
　　debutsi
$$t[i] \leftarrow \frac{T_r[i]}{T_{max}}$$
$$c[i] \leftarrow \frac{C_r[i]}{C_{max}}$$
　　Som_coutTemps ← Som_coutTemps + (αc[i] + (1-α)t[i])
　　somme[i] ← Som _qlte - Som_coutTemps
　finsi
Finpour
Fin

*Fig 5 : code de la procedudureevaluation*

### 4.3.3. Algorithme Optimisation

*L'algorithmeevaluation* nous a permis de sélectionner un ensemble de composantsrépondant aux besoins de qualité définis par l'utilisateur. A l'aide de la procédure optimisation, nous allons trier ce vecteur de réels désignant le degré de qualité des composants retenus.

**Procedure Optimisation** ($M_{ax}$ : entier)

**Entrée** :i,j   entier
somme  tableau [1.. $M_{ax}$] de réels
**Sortie** : somme  tableau [1.. $M_{ax}$] de réels triés par ordre décroissant
　debut
　　pour *i* de 1 à $M_{ax}$ -1 faire
　　　debut
　　　　pour *j* de *i* +1 à $M_{ax}$ faire
　　　　　debut
*si*   (somme *[i ]* >somme*[j]* )   *alors*
　　　　　　debutsi
*tampon* = somme *[i ]*

ARIMA - Phase de Validation

somme *[i ]*= somme *[j ]*
        somme*[j]=tampon*
*Finsi*
    *Finpour*
  *Finpour*
*Fin*

*Fig 6 : code de la procédure optimisation*

## 5.  Conclusion

Notre travail repose sur trois approches. Nous avons construit un modèle de qualité qui prend en compte les caractéristiques définissant la fiabilité  et la sûreté du logiciel à construire. Ensuite nous construit un modèle de sélection de composants logiciels dans une bibliothèque ou sur le net. Enfin nous avons défini une métrique qui prend en compte le coût financier et le temps d'adaptation  des composants sélectionnés. Cette métrique évalue et simule la qualité du composant en optimisant les paramètres dont le coût financier et le temps.
Cette approche est soutenue par l'algorithme SelectCompo construit.
Plusieurs aspects restent à développer. Il s'agit de prendre en compte la sélection de composants logiciels dans diverses bibliothèques pour toute plateforme. Cela permettra de résoudre le problème d'interopérabilité de ces composants sur différentes plateformes. Dans les futurs travaux, nous pourrons tester la facilité d'intégration de ces composants sur une plateforme donnée.
Afin de résoudre les difficultés de déploiement de certains composants logiciels sur une combinaison de plateforme, de fourniture de service à des utilisateurs ou des clients anonymes, nous pourrions orienter nos recherches  vers les web services. Cela aura pour avantage de réagir rapidement à tous changement en s'assurant de la fiabilité et la de sécurité de ces composants réutilisables.

## Bibliographie